\documentclass{elsart5p}
\usepackage{graphics}
\usepackage{graphicx}
\usepackage{epsfig}
\usepackage{amssymb}
\usepackage{amsmath}
\usepackage{color}

\newcommand{\dd}{\mbox{$\textrm{d}$}}
\def\fmn#1#2{\mbox{${\textstyle \frac{#1}{#2}}$}}
\begin{document}
\begin{frontmatter}

\title{Coherent pion production in proton-deuteron collisions}
\author[ikp,dubna]{S.~Dymov\corauthref{cor1}},
\ead{s.dymov@fz-juelich.de}
\author[dubna]{V.~Shmakova},
\author[ikp,tbilisi]{D.~Mchedlishvili},
\author[dubna]{T.~Azaryan},
\author[gatchina]{S.~Barsov},
\author[gatchina]{A.~Dzyuba},
\author[ikp]{R.~Engels},
\author[ikp]{R.~Gebel},
\author[Muenster]{P.~Goslawski},
\author[Lanzhou]{B.~Gou},
\author[gatchina,Aachen]{K.~Grigoryev},
\author[ikp]{M.~Hartmann},
\author[ikp]{A.~Kacharava},
\author[dubna]{V.~Komarov},
\author[Muenster]{A.~Khoukaz},
\author[cracow]{P.~Kulessa},
\author[dubna]{A.~Kulikov},
\author[dubna]{V.~Kurbatov},
\author[tbilisi]{N.~Lomidze},
\author[ikp]{B.~Lorentz},
\author[tbilisi]{G.~Macharashvili},
\author[ikp]{S.~Merzliakov},
\author[Muenster]{M.~Mielke},
\author[ikp,gatchina]{S.~Mikirtytchiants},
\author[tbilisi]{M.~Nioradze},
\author[ikp]{H.~Ohm},
\author[Muenster]{M.~Papenbrock},
\author[ikp]{D.~Prasuhn},
\author[ikp]{F.~Rathmann},
\author[ikp]{V.~Serdyuk},
\author[ikp]{H.~Seyfarth},
\author[ikp]{H.~Str\"oher},
\author[tbilisi]{M.~Tabidze},
\author[ikp,MSU]{S.~Trusov},
\author[dubna]{D.~Tsirkov},
\author[dubna,MSU-Uzikov,Dubna-Uni]{Yu.~Uzikov},
\author[gatchina,Bonn]{Yu.~Valdau},
\author[ucl]{C.~Wilkin}, and
\author[ZEL]{P.~W\"ustner}
\corauth[cor1]{Corresponding author.}

\address[ikp]{Institut f\"ur Kernphysik, Forschungszentrum J\"ulich, D-52425 J\"ulich, Germany}
\address[dubna]{Laboratory of Nuclear Problems, Joint Institute for Nuclear
  Research, RU-141980 Dubna, Russia}
\address[tbilisi]{High Energy Physics Institute, Tbilisi State University, GE-0186
Tbilisi, Georgia}
\address[gatchina]{St. Petersburg Nuclear Physics Institute, RU-188350 Gatchina,
  Russia}
\address[Muenster]{Institut f\"ur Kernphysik, Universit\"at M\"unster,
D-48149 M\"unster, Germany}
\address[Lanzhou]{Institute of Modern Physics, Chinese Academy of Sciences, Lanzhou 730000, China}
\address[Aachen]{Physikalisches Institut (IA), RWTH Aachen, 52056 Aachen, Germany}
\address[cracow]{Institute of Nuclear Physics, PL-31342 Cracow,
Poland}
\address[MSU]{Skobeltsyn Institute of Nuclear Physics, Lomonosov Moscow State University, RU-119991 Moscow, Russia}
\address[MSU-Uzikov]{Department of Physics, Lomonosov Moscow State   University, RU-119991 Moscow, Russia}
\address[Dubna-Uni]{ Dubna State University, RU-141980 Dubna, Russia}
\address[Bonn]{Helmholtz-Institut f\"ur Strahlen- und Kernphysik, Universit\"at Bonn, Bonn, Germany}
\address[ucl]{Physics and Astronomy Department, UCL, Gower Street, London WC1E 6BT, United Kingdom}
\address[ZEL]{Zentralinstitut f\"ur Elektronik, Forschungszentrum J\"ulich, D-52425 J\"ulich, Germany}

\date{\today}

\begin{abstract}
Values of the proton analysing power in the
$pd\to{}^{3}\textrm{He}\,\pi^0/^{3}\textrm{H}\,\pi^+$ reactions at
350-360~MeV per nucleon were obtained by using a polarised proton beam
incident on a deuterium cluster-jet target and with a polarised deuteron beam
incident on a target cell filled with polarised hydrogen. These results have
a much larger angular coverage than existing data. First measurements are
also presented of the deuteron vector analysing power and the deuteron-proton
spin correlations. Data were also obtained on the deuteron-proton spin
correlation and proton analysing power at small angles at 600~MeV per
nucleon, though the angular coverage at this energy was much more restricted
even when using a deuteron beam. By combining the extrapolated values of the
spin correlations to the forward or backward directions with published
measurements of the deuteron tensor analysing powers, the relative phases
between the two non-vanishing amplitudes were evaluated.
\end{abstract}

\end{frontmatter}
%
%

The simplest coherent pion production reaction, where all the final nucleons are
bound in a nucleus, is $pp\to d\pi^+$ and the associated literature is very
extensive and the database enormous~\cite{ARN1993}. However, because of the
identical nature of the two initial protons, odd and even pion waves do not
interfere in the differential cross section which, as a consequence, is
symmetric around $90^{\circ}$ in the centre-of-mass (c.m.) frame. The first
more general coherent pion production reaction is
$pd\to{}^{3}\textrm{He}\,\pi^0$, where striking effects arising from the
interference between $s$ and $p$ partial waves are observed in the
differential cross section even very near threshold~\cite{PIC1992,NIK1996}.

Since there is only one isospin amplitude, the cross section for
$pd\to{}^{3}\textrm{H}\,\pi^+$ should be twice that for
\mbox{$pd\to{}^{3}\textrm{He}\,\pi^0$} but all the polarisation observables
should be identical for the two reactions. On the other hand, the
$\half^+1^+\to \half^+0^-$ spin structure leads to six independent
amplitudes, all of which are functions of the pion production angle. There
are therefore a very wide range of possible experiments, which have been
discussed in detail by Uzikov~\cite{UZI2008}. The observables that are
accessible at the COSY accelerator involve the polarisations of the incident
particles and measurements have been made of the proton and deuteron vector
analysing powers and the proton-deuteron vector spin correlations.

Measurements of the differential cross section and proton analysing power for
$\pol{p}d\to{}^{3}\textrm{He}\,\pi^0$ were undertaken at TRIUMF at
350~MeV~\cite{CAM1987}, though the acceptance of their spectrometer was very
limited near the forward and backward directions. The cross section data show
a steep but rather featureless drop from the forward (small c.m.\ angle
$\theta_{\pi}^{cm}$ between the proton and the pion) to the backward
directions and this was confirmed by later measurements by the GEM
collaboration~\cite{BET2001,ABD2003}. Much more structure is, however, seen
in the distribution of the corresponding analysing power~\cite{CAM1987}. In
addition to improving significantly the angular coverage of these TRIUMF
data, we show for the first time data on the deuteron-proton vector spin
correlations and the deuteron vector analysing power. It is hoped that these
new observables will stimulate further theoretical work.

The data that we report here came as by-products of measurements of
quasi-free pion production in proton-neutron collisions using a deuterium
target~\cite{DYM2012} and deuteron beam~\cite{DYM2013} in the region of
353--363~MeV per nucleon and also deuteron charge exchange at 600~MeV per
nucleon~\cite{MCH2013}. The experimental conditions, including the crucial
determination of the beam and target polarisations, were already described in
these publications so that we can here be relatively brief.

The experiments were carried out at the ANKE spectrometer
facility~\cite{BAR1997}, which is installed inside the COSY cooler
synchrotron storage ring of the Forschungszentrum-J\"{u}lich. The proton
analysing power was first studied by using a polarised 353~MeV proton beam
incident on a deuterium cluster-jet target~\cite{KHO1999}. The polarisation
of the circulating proton beam, $|p|=0.66\pm 0.06$, was reversed in direction
every six minutes.

The triton or $^3$He from the
$pd\to{}^{3}\textrm{H}\,\pi^+/^{3}\textrm{He}\,\pi^0$ reaction was registered
in the ANKE Forward Detector (FD). The FD comprises a set of multiwire
proportional and drift chambers and a two-plane scintillation
hodoscope~\cite{DYM2004}. The data were collected with a dedicated trigger
for high energy losses in one of the counters of the first hodoscope plane.
$^3$He and tritons were then selected using the calibrated energy loss in the
hodoscope and the particle momentum, which was reconstructed from the MWPC
information. All $^3$He and low energy tritons, which fly backwards in c.m.\
frame, stopped in the first plane of the hodoscope. However, the fast
(forward-going) tritons reached the second plane and in this case the time of
flight between the planes was used as an additional criterion for the
particle identification. The pion was finally identified through the missing
mass in the reaction.

Since 353~MeV is far above the threshold for the
$pd\to{}^{3}\textrm{H}\,\pi^+/^{3}\textrm{He}\,\pi^0$ reactions, there is no
acceptance in ANKE for events in the central region of c.m.\ angles.
Nevertheless the angular coverage was maximised by combining the
spin-dependent data associated with fast $^3$He and both fast and slow $^3$H.

Isospin invariance requires that the ANKE results for the proton analysing
power should be identical for $^3$H and $^3$He detection and these data are
compared with the TRIUMF values~\cite{CAM1987} in terms of the c.m.\ angle
$\theta_{\pi}^{cm}$ between the initial proton and final pion. Though the
data sets are consistent, the ANKE results define the behaviour for small and
large $\theta_{\pi}^{cm}$ more clearly and with much higher statistics.
Despite the smooth behaviour of the differential cross sections with angle,
the rich structure in $A_y^p$ indicates that many partial waves with
different phases contribute actively at this energy. The general behaviour of
$A_y^p$ is also confirmed by ANKE measurements with a polarised hydrogen
cell, to which we now turn.

\begin{figure}[hbt]
\begin{center}
\includegraphics[width=1.0\columnwidth]{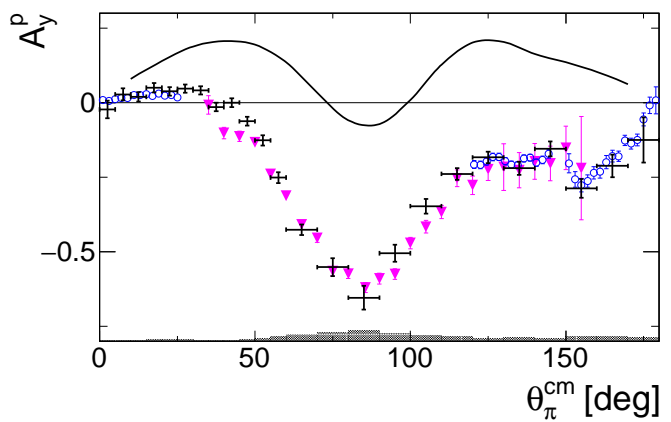}
\end{center}
\caption{TRIUMF data on the proton analysing power $A_y^p$ in the
$\pol{p}d\to{}^{3}\textrm{He}\,\pi^0$ reaction at 350~MeV~\cite{CAM1987}
(magenta triangles) are compared to ANKE results at 353~MeV (blue open
circles) obtained by detecting fast $^3$He ($\theta_{\pi}^{cm}\gtrsim
150^{\circ}$) and $^3$H ($120^{\circ}\lesssim\theta_{\pi}^{cm}\lesssim
145^{\circ}$), and slow $^3$H ($\theta_{\pi}^{cm}\lesssim 25^{\circ}$). We do
not include here the 9\% systematic error that is mainly associated with
uncertainties in the COSY beam polarisation. Also shown are the ANKE values
at 363~MeV per nucleon deduced from measurements with a deuteron beam
incident on a long cell filled with polarised hydrogen gas (black crosses).
The shaded area indicates the systematic uncertainties in the deuteron beam
measurements. There is no evidence for any violation of isospin invariance,
which requires that the analysing powers measured with $^3$He and $^3$He
detection should be identical. The curve corresponds to the predictions by
Falk in a cluster-model approach~\cite{FAL2000,FAL2010}.} \label{fig:ayp}
\end{figure}

In the experiment with the deuteron beam at an energy of 726~MeV, the forward
boost coming from the higher beam momentum means that all the $^3$He and
$^3$H reached the second layer of the FD and so the timing information
associated with hits in the first and second layers of this detector is even
more critical. The forward boost also increases significantly the angular
acceptance. Only vector polarisation modes of the deuteron source were used
and these had ideal values of $p_{\uparrow}=\frac{2}{3}$ and
$p_{\downarrow}=-\frac{2}{3}$ in the $y$ direction\footnote{We are here using
the notation where $\hat{z}$ lies along the beam direction, $\hat{y}$
represents the upward normal to the plane of the COSY ring, and the other
transverse direction $\hat{x}$ lies along $\hat{y}\times\hat{z}$.}. However,
such high figures are never achieved in practice and the measured values had
magnitude $|p|= 0.50\pm 0.05$, with the tensor polarisation being below 2\%
for both modes. The target was a 39~cm long Teflon-coated aluminum storage
cell fed from an atomic beam source. The orientation of the hydrogen spin was
reversed every five seconds and the mean target polarisation was determined
from the $\pol{p}\,\pol{n}\to d\pi^0$ calibration reaction to be $|q| = 0.69
\pm 0.04$~\cite{DYM2013}.

The experimental conditions are clearly not as clean for the storage cell
compared to the cluster-jet target. There is a background from events arising
from the aluminium walls and the target is far from being point-like.
Nevertheless, having both beam and target polarised it is possible to extract
proton and deuteron analysing powers as well as the spin correlations. The
values obtained for $A_y^p$ from the deuteron beam experiment, which are also
shown in Fig.~\ref{fig:ayp}, also cover the central region of angles. Though
the statistics are limited and the angular bins wider, the resulting data are
completely consistent with both the TRIUMF and the ANKE cluster-jet data
taken in $pd$ kinematics.

\begin{figure}[hbt]
\begin{center}
\includegraphics[width=1.0\columnwidth, angle=0]{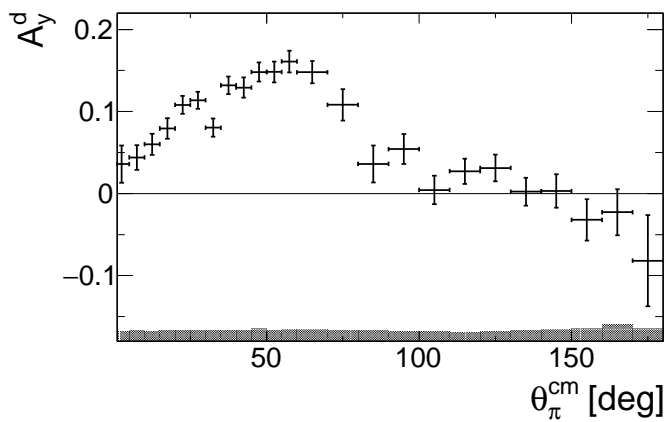}
\end{center}
\caption{Deuteron vector analysing power $A_y^d$ measured in the
$\pol{d}p\to{}^{3}\textrm{He}\,\pi^0/^{3}\textrm{H}\,\pi^+$ reactions at
726~MeV. The data are presented in terms of the c.m.\ angle between the
proton and pion. The shaded area indicates the systematic uncertainties in
the measurement. } \label{fig:ayd}
\end{figure}

Since the tensor polarisation of the deuteron beam is vanishingly small, the
deuteron vector analysing power $A_y^d$ can also be extracted from the data
by looking at the dependence of the counting rates on the vector polarisation
of the beam. The results obtained from the combined
$\pol{d}p\to{}^{3}\textrm{He}\,\pi^0/^{3}\textrm{H}\,\pi^+$ data are shown in
Fig.~\ref{fig:ayd}. The error bars here are somewhat larger than those for
the corresponding $A_y^p$ data of Fig.~\ref{fig:ayp} due to the choice of the
beam polarisation modes. It is perhaps significant that the abrupt change in
$A_y^d$ occurs at $\theta_{\pi}^{cm}\approx 80^{\circ}$, which is close to the deep
minimum in $A_y^p$.

\begin{figure}[htb]
\centering
\includegraphics[width=0.9\columnwidth]{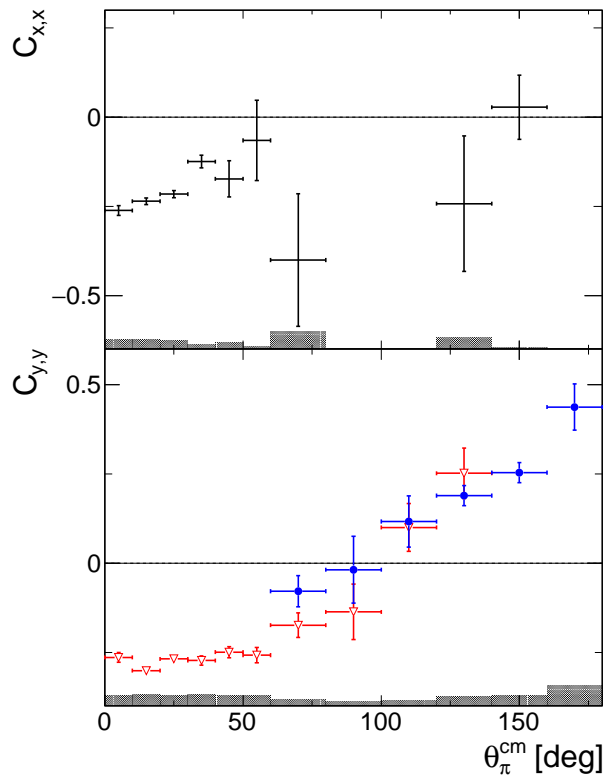}
\caption{Transverse spin correlation coefficients $C_{x,x}$ and $C_{y,y}$ in
the $\pol{d}\pol{p}\to {}^3\rm{He}\,\pi^0$  and $\pol{d}\pol{p}\to
{}^3\rm{H}\,\pi^+$  reactions at 363~MeV per nucleon. In the $C_{y,y}$ case
the (red) inverted triangles were obtained through $^3$He detection and the
(blue) circles through $^3$H detection. The shaded area indicates the
systematic uncertainties in the measurement.
} \label{fig:CxxCyy}
\end{figure}

The acceptance in the COSY experiments is limited by the size of the vertical
gap in the ANKE analysing magnet, which results in the $C_{y,y}$ coefficient
being measured over a much wider angular range than $C_{x,x}$. The
consequences of this are illustrated in Fig.~\ref{fig:CxxCyy} where, by
considering both $^{3}\textrm{He}\,\pi^0$ and $^{3}\textrm{H}\,\pi^+$
detection, values of $C_{y,y}$ could be obtained over the whole angular range
at 726~MeV, whereas the $C_{x,x}$ measurements were only possible for small
angles close to the forward and backward directions. It should be noted here
that $C_{y,y}$ changes sign around $90^{\circ}$.

The main difference between the double-polarisation measurements at 726 and
1200~MeV was the choice of the beam polarisation modes. In the latter case,
in addition to an unpolarised mode, these included a pure vector polarised
mode with ideal values of $(p_z,p_{zz})=\left(-\fmn{2}{3},0\right)$. This
experiment was carried out over two separate beam times, with the measured
values of the beam polarisations being $p_z=-0.53\pm 0.05$ and $p_z=-0.62\pm
0.08$, and average target polarisations of $|q|=0.66\pm 0.03$ and
$|q|=0.78\pm 0.03$.

\begin{figure}[htb]
\centering
\includegraphics[width=1.0\columnwidth]{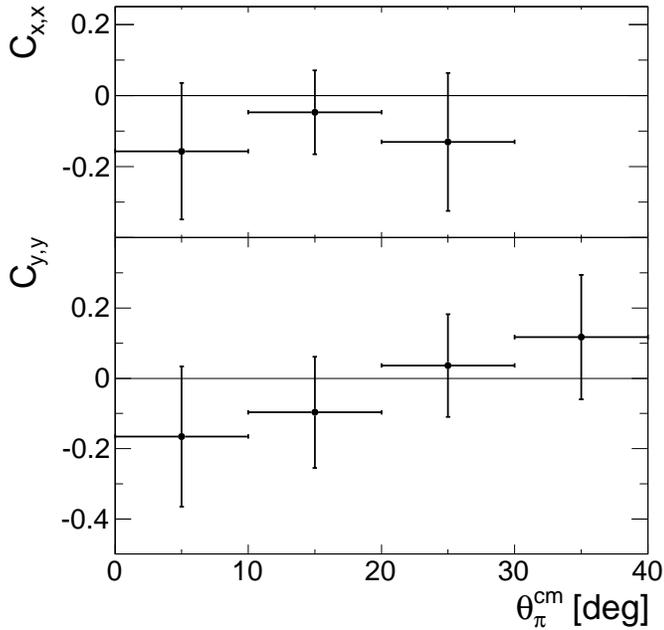}
\caption{Transverse spin correlation coefficients $C_{y,y}$
and $C_{x,x}$
 measured in the $\pol{d}\pol{p}\to
{}^3\rm{He}\,\pi^0$ reaction at 600~MeV per nucleon. The systematic errors
are below $0.03$.} \label{fig:C600}
\end{figure}

At 600~MeV per nucleon it is no longer feasible to make a clean selection of
tritons by their energy loss in the scintillation hodoscope of the FD. As a
consequence, only data on the $\pol{d}\pol{p}\to {}^3\rm{He}\,\pi^0$ reaction
are shown at this higher energy and this limits severely the angular range
covered in the experiment. The $C_{x,x}$ and $C_{y,y}$ coefficients measured
at small angles are presented in Fig.~\ref{fig:C600}, though the error bars
are much more significant than at 363~MeV per nucleon.

\begin{figure}[hbt]
\begin{center}
\includegraphics[width=1.0\columnwidth]{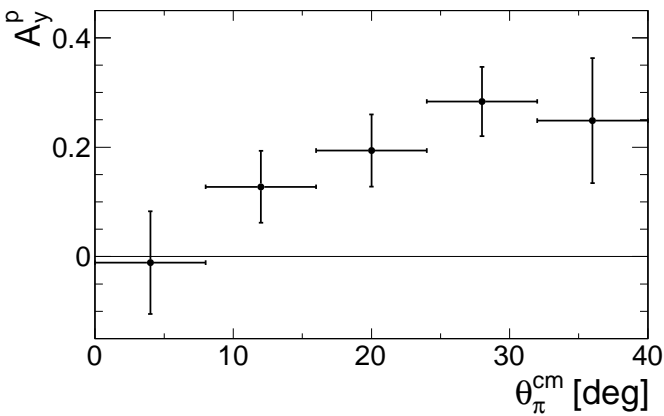}
\end{center}
\caption{The proton analysing power $A_y^p$ for the $d\pol{p}\to
^3$He$\,\pi^0$ reaction at 600~MeV per nucleon extracted from data obtained
with a polarised hydrogen target. Systematic uncertainties, which were
dominated by those in the target polarisation, were below 5\%.}
\label{fig:ayp600}
\end{figure}

The uncertainties in the corresponding deuteron vector analysing powers are
very large and these data are not shown. However, values of the proton
analysing powers could be extracted for small angles by exploiting the
polarisation of the hydrogen in the target and these data are presented in
Fig.~\ref{fig:ayp600}. These results for $\theta_{\pi}^{cm} < 40^{\circ}$ are
significantly larger than those of the lower energy data shown in
Fig.~\ref{fig:ayp}.

In the forward and backward directions the number of independent amplitudes
reduces from six to two and these may be written as~\cite{GER1990}
\begin{equation}
 F(dp \to{} ^{3\!}\rm{He} \pi^0) = \overline{u}_\tau\,\vec{p}\!\cdot\!(A\vec{\epsilon}+iB\vec{\epsilon}\times\vec{\sigma})u_p.
\label{eq:1}
\end{equation}
Here $\vec{\epsilon}$ is the deuteron polarisation vector, $\vec{p}$ the
proton c.m.\ momentum, and $u_p$ and $u_\tau$ are the initial and final
fermion spinors. Apart from one discrete ambiguity, all the possible
experimental information is contained in the initial-state spin observables:
\begin{eqnarray}
\nonumber
\hspace{2.5cm}\frac{\dd\sigma}{\dd\Omega} & = & \frac{kp}{3}(|A|^2+2|B|^2), \label{eq:2}\\
\nonumber
T_{20} & =& \sqrt{2}\frac{|B|^2-|A|^2}{|A|^2+2|B|^2}, \label{eq:3}\\
C_{y,y}=C_{x,x}& = & -\frac{2Re(A^*B)}{|A|^2+2|B|^2}, \label{eq:4}
\end{eqnarray}
where $k$ is the pion c.m.\ momentum. The forms of these equations are
identical for $\theta_{\pi}^{cm}=0^{\circ}$ and $180^{\circ}$ though, of
course, the amplitudes $A$ and $B$ are different in the two cases.

The deuteron tensor analysing power $T_{20}$ was measured in the
forward/backward directions at Saclay~\cite{KER1986} and the values
interpolated at 726~MeV are $T_{20}(0^{\circ})=-1.01\pm0.01$ and
$T_{20}(180^{\circ})=-1.10\pm0.06$. However, it is easily shown from
Eq.~(\ref{eq:4}) that the corresponding spin correlation in the forward or
backward direction is bounded by
\begin{equation}
\label{bound}
\left(C_{y,y}\right)^2 \leq \fmn{4}{9}\left[1-T_{20}/\sqrt{2}-(T_{20})^2\right],
\end{equation}
from which one sees quite generally that $|C_{y,y}|\leq 1/\sqrt{2}$.

Using the forward value of $T_{20}$ measured at Saclay, Eq.~(\ref{bound})
shows that $|C_{y,y}(0^{\circ})|\leq 0.56\pm0.01$ to be compared to the value
$-0.27\pm0.03$ deduced from extrapolating the combined data of
Fig.~\ref{fig:CxxCyy} to the forward direction. The error bars are larger in
the backward direction where one finds from the data of Fig.~\ref{fig:CxxCyy}
that $C_{y,y}(180^{\circ})=+0.46\pm0.04$ compared to the upper bound from
Eq.~(\ref{bound}) of $0.50\pm0.04$\footnote{All the error bars in the
extrapolated values of $C_{y,y}$ or the phase angle $\phi$ include the
systematic uncertainties.}.

The change in sign of $C_{y,y}$ between the backward and forward directions
is significant. If the phase angle is defined by $\phi=\arg{(B/A)}$ then, in
the forward direction, $\cos\phi = 0.49\pm 0.05$ whereas in the backward
direction $\cos\phi = -0.90\pm 0.10$\footnote{There remains a discrete
ambiguity corresponding to the sign of $\sin\phi$.}. The change in sign of
$Re(A^*B)$ could be due to structure in either amplitude. The Saclay data
indicate that this is likely to be caused by the $B$ amplitude because
$B(180^{\circ})$ seems to have a zero in the vicinity of
$T_d=650$~MeV~\cite{KER1986}.

The deuteron tensor analysing power bound of Eq.~(\ref{bound}) provides
little real constraint at 1.2~GeV. Using the Saclay value of
$T_{20}(0^{\circ})=-0.66\pm0.02$ one finds that
$|C_{y,y}(0^{\circ})|<0.68\pm0.01$ to be compared with the extrapolated value
of Fig.~\ref{fig:C600}, $C_{y,y}(0^{\circ}) =-0.10\pm0.08$. These mean that
in the forward direction the $A$ and $B$ amplitudes at 600~MeV per nucleon
are almost completely out of phase, with $\cos\phi=0.14\pm 0.12$.

The simplest phenomenological model used to describe the $pd\to {}^3\rm{He}
\,\pi^0$ reaction is the cluster approach, first proposed by
Ruderman~\cite{RUD1952}. It is here assumed that the $pn\to d\pi^0$ reaction
takes place on the neutron in the target deuteron and that the deuteron
produced is captured by the spectator proton to generate the observed $^3$He.
Due to the mass differences, there are ambiguities in the implementation of
the kinematics in any such model. The \textit{prescription} employed by
Falk~\cite{FAL1994} assumes that the c.m.\ momentum of the pion is the same
in the $pd\to {}^3\rm{He} \,\pi^0$ and $pn\to d\pi^0$ reactions. Though this
model can describe the data taken very near threshold~\cite{NIK1996}, its
predictions for the proton analysing power at 363~MeV in Fig.~\ref{fig:ayp}
are not encouraging~\cite{FAL1994}. The model should work best at small
$\theta_{\pi}^{cm}$ and the sign of $A_y^p$ is correct there but the magnitude is much
too large. The predicted dip around $90^{\circ}$ is seen in the data but the
large angle prediction is even wrong in sign as well as magnitude. The proton
analysing power depends sensitively upon the relative phases of amplitudes,
which might be changed by secondary effects. ``It is clear that the model is
lacking in one or more aspects; one of these might well be the neglect of the
$\{pp\}$ singlet contribution''~\cite{FAL2010}.

In their implementation of the cluster model, Germond and Wilkin attempted to
include singlet contributions from the $pp\to \{pp\}\pi^0$ in addition to the
dominant $pn\to d\pi^0$~\cite{GER1990}. They employed a slightly different
kinematic \emph{prescription} to Falk, where the pion momenta in the
$\pi^0\,{}^3\textrm{He}\to pd$ and $\pi^0 d \to pn$ were assumed to be
identical in the laboratory frame. However, they only carried out the
calculations in the forward/backward directions and so no estimates could be
made of $A_y^p$ or $A_y^d$ and, moreover, no predictions were made of the
spin-correlation parameter. Furthermore, the signs of the $D$-wave
contributions have also been questioned~\cite{KOB2013}.

It was shown that the inclusion of the spin-singlet terms was crucial for the
predictions of both the cross section and the deuteron tensor analysing
power. Reasonable values could then be obtained in the backward direction up
to $T_d\approx 700$~MeV, though the range of validity is somewhat larger in
the forward direction. Since the input that generates the $A$ and $B$
amplitudes are essentially independent their relative phase $\phi$ must
depend upon the details of the model. However, the change in the phase of $B$
between the forward and backward directions seems to be of more general
interest.

In summary, we have measured the proton analysing power in the
$pd\to{}^{3}\textrm{He}\,\pi^0$ and $pd\to{}^{3}\textrm{H}\,\pi^+$ reactions
with a polarised proton beam incident on a deuterium cluster-jet target and,
in inverse kinematics, with a polarised deuteron beam incident on a target
cell filled with polarised hydrogen. These results, obtained at 350-360~MeV
per nucleon, led to values of the proton analysing power that were consistent
with the TRIUMF measurements though with much larger angular coverage. This
was facilitated by invoking isospin invariance, which requires that all
analysing powers and other spin observables should be identical for $^3$He
and $^3$H detection.
\pagebreak%

Of even greater importance for the theoretical modeling are the first ever
measurements of the deuteron vector analysing power and the deuteron-proton
vector spin correlations. These were obtained at 363~MeV per nucleon but
small angle spin correlations and proton analysing powers were also extracted
at 600~MeV per nucleon from data taken with a target cell filled with
polarised hydrogen. The combination of Saclay $T_{20}$ and COSY $C_{y,y}$
data shows that the two non-vanishing amplitudes are almost completely out of
phase in the forward direction at 600~MeV per nucleon. However, in the
350-360~MeV per nucleon region, there is a large overlap in phase in both the
forward and backward directions, though the relative phase $\phi$ changes
significantly between these two extremes.
\\

We are grateful to the COSY crew for providing such good working conditions,
especially of the polarised beams. We thank W.R.~Falk for correspondence and
for providing numerical values of the $A_y^p$ predictions used in
Fig.~\ref{fig:ayp}. This work has been partially supported by the COSY
FFE programme and the Shota Rustaveli National Science Foundation.

%
%

%
\end{document}